\documentstyle[12pt]{article}
\def\be {\begin{eqnarray}}
\def\ee {\end{eqnarray}}
\def\beq {\begin{equation}}
\def\eeq {\end{equation}}
\def\bi {\begin{itemize}}
\def\ei {\end{itemize}}
\def\ben {\begin{enumerate}}
\def\een {\end{enumerate}}
\def\ni {\noindent}
\def\sutwo {$SU(2)_L \times SU(2)_R$ }
\def\suthree {$SU(3)_L \times SU(3)_R$ }
\def\del {\partial}

\def\om3 {{\cal O}(m_\pi^3)}
\newcommand{\nno}{\nonumber}

\newcommand\half{{1\over 2}}
\newcommand\third{{1\over 3}}

\newcommand{\Mpi}{m_\pi}

\newcommand{\Zeff}{Z_{\rm eff}}

\begin{document}
\setcounter{page}{1}
\hfill {NORDITA-95/7 N}
\vskip -0.1cm
\hfill{\tt nucl-th/9502003}
%\hfill\today

\vskip 1cm

\begin{center}
{\large\bf S-wave Meson-Nucleon Interactions and the\\}
{\large\bf Meson Mass in Nuclear Matter from\\}
{\large\bf Chiral Effective Lagrangians\\}
\end{center}

\vskip 1cm

\begin{center}
{\large Vesteinn Thorsson$^{a,}$\footnote{Email: thorsson{\verb+@+}nordita.dk}
and Andreas Wirzba$^{b,}$\footnote{Email:
wirzba{\verb+@+}crunch.ikp.physik.th-darmstadt.de}}\\[4mm]
{\it $^{a}$ NORDITA, Blegdamsvej 17, DK--2100 Copenhagen \O, Denmark\\[2mm]
$^{b}$ Institut f\"ur Kernphysik, Technische Hochschule Darmstadt,\\
Schlo{\ss}gartenstra{\ss}e 9, D--64289 Darmstadt, Germany }
\end{center}

\vskip 2cm
\centerline{\bf Abstract}

\noindent

Chiral effective lagrangians may differ in their prediction
of meson-nucleon scattering amplitudes off-meson-mass-shell,
but must yield identical S-matrix elements.
We argue that the effective meson mass in nuclear matter
obtained from chiral effective lagrangians is also unique.
Off-mass-shell amplitudes obtained using the PCAC
choice of pion field must therefore not  be viewed as
fundamental constraints
on the dynamics, the determination of the effective meson mass
in nuclear matter or the possible existence of meson condensates
in the ground state of nuclear matter.
This hypothesis is borne out by a calculation of the effective
mass in two commonly employed
formulations of chiral perturbation theory which yield different
meson-nucleon scattering amplitudes off-meson-mass-shell.

\vfill
%\noindent nucl-th/xxxxxxx

\newpage

%%%%%%%%%%%%%%%%%%%%%%%%%%%%%%%%%%%%%%%%%%%%%%%%%%%%%%%%%%%%%%%%
%
%	Section 1
%
%%%%%%%%%%%%%%%%%%%%%%%%%%%%%%%%%%%%%%%%%%%%%%%%%%%%%%%%%%%%%%%%
\section{Introduction}

\setcounter{equation}{0}
\indent

In recent years, there has been considerable discussion of
whether Bose condensates of charged mesons may be found  in dense
nuclear matter, such as that formed in collisions of heavy
nuclei, in cores of collapsing stars, or in the interior
of neutron stars.
These studies have been motivated by the suggestion of Kaplan and
Nelson\cite{kapnel} that attractive S-wave interactions
between kaons and nucleons could lower the effective mass
of kaons to the extent that kaons could condense in dense neutron star
matter,
at several times nuclear saturation density.
The kaon-nucleon interactions of Ref.\cite{kapnel} are obtained
from a flavour \suthree lagrangian, originally due to
Manohar and Georgi\cite{mg}, but with the inclusion of
additional terms in the expansion in powers of $\Lambda^{-1}$,
where $\Lambda \simeq 1$ GeV is the scale of chiral symmetry breaking.
The S-wave attraction is driven by two terms in the lagrangian,
one related to the exchange
of vector mesons between kaons and nucleons and the other
proportional to the kaon-nucleon sigma term, $\Sigma^{KN}$,
corresponding to the exchange of a scalar meson.
The value of the critical density for kaon condensation in such
models varies, due to the large uncertainty for estimates of the
sigma term: $\Sigma^{KN} \approx (200-400)$ MeV.
In the dense interior of neutron stars, a condensate may be formed
if the effective kaon mass is brought down to the value of the charge
chemical potential, typically around 200 MeV at the densities
under consideration,
converting the electrons and muons in beta equilibrium with
the matter to kaons.
For the range of $\Sigma^{KN}$ quoted above, the critical
density lies in the range 3-5~$\rho_0$, in terms of the
nuclear matter saturation density $\rho_0 =  0.16 \,\,{\rm fm}^{-3}$
\cite{kapnel,bkrt,tpl}.

The formation of a condensate depends crucially on the nature of
the kaon-nucleon interaction.
A number of extensions of the original model\cite{kapnel} have
been made.
One direction of development has been the inclusion of additional
terms in the chiral expansion. The standard chiral counting orders
scattering amplitudes as a power series in the characteristic
energy-momentum scale, $Q$, with the leading order kaon-nucleon
amplitude
being ${\cal O}(Q^1)$.
Brown {\it et al.}\cite{blrt}
included the complete set of terms
of the same order as the sigma term,
${\cal O}(Q^2)$,
and fitted the expansion coefficients to $K N$
scattering lengths.  Lee and
collaborators\cite{chlee} have included the
next order in chiral perturbation theory,   ${\cal O}(Q^3)$,
which includes one-loop diagrams, and extended the fit to include
${\bar K}N$ scattering amplitudes.
Several studies have also included contributions difficult
to access in chiral perturbation theory, such as
resonance contributions\cite{chlee} and effects
from nuclear correlations\cite{chlee,ppt}.
Another line of development,
suggested by Brown, Koch and Rho\cite{bkr}
has been the application of the ideas of Ref.\cite{kapnel}
to S-wave pion condensation.  However, the corresponding pion-nucleon
sigma term $\Sigma^{\pi N} \simeq 45$ MeV
is considerably smaller than
$\Sigma^{KN}$.  Also, interactions mediated by the exchange of
vector mesons, $\rho$-meson exchange in the case of pions, and
both $\rho$- and $\omega$- meson exchange in the case of kaons,
are generally favourable for S-wave kaon condensation, but not for
S-wave pion condensation.  More specifically,
the pion-proton vector meson exchange potential,
is only half as attractive as that for kaons, while the corresponding
pion-neutron potential is repulsive,
and of the same magnitude as for kaons.
The vector mediated pion-nucleon interaction is
thus repulsive in neutron-rich matter.
The inclusion of additional terms in the chiral expansion,
fitted to pion-nucleon scattering lengths, therefore leads to
a in-medium
pion mass that increases slightly with density. The increase
to linear order in density was reemphasized in the
work of Delorme, Ericson and Ericson\cite{dee}.

Recently, several author have
claimed that the method used to motivate
and describe meson condensation is incorrect\cite{dee,sc1,sc2}.
They argued   that chiral effective lagrangians are inconsistent with
current algebra and PCAC\cite{dee,sc1} for the following reason:
off-meson-mass-shell meson-nucleon scattering amplitudes
computed from the lagrangian conventionally employed in the
description of meson condensation (i.e.\
of the  Kaplan and Nelson\cite{kapnel}
type) are not those obtained using PCAC.
Secondly, they claimed
that the incorporation of these
off-meson-mass-shell amplitudes in the calculation inevitably
leads to an effective repulsion which serves to inhibit meson
condensation\cite{sc1,sc2}.

In this Letter, we argue -- in line with well-established
theorems\cite{ccwz} --
that the S-wave meson-nucleon scattering
amplitudes obtained off-meson-mass-shell are subject to
the choice of meson field, are therefore entirely unphysical, and
are thus not to be viewed as constraints on a theory.
However, the effective meson mass in nuclear matter
behaves as the on-meson-mass-shell
scattering amplitude, in that it is
{\it independent} of the choice made for the meson field.
(~The identical conclusion was reached independently
in the most recent work of Lee et al.\cite{chlee}, through slightly
different arguments.~)
We argue that this is to be understood as a consequence of a general
rule:  any physically relevant observable is independent of
the choice of meson field variables, as is the case for S-matrix
elements\cite{ccwz}.
To support this conjecture,
using a formulation of chiral perturbation theory
for which the canonical meson field {\it is} to be identified with
the divergence of the axial vector current, namely that originating
in the work of Gasser and Leutwyler\cite{gl},
we obtain an effective meson
mass identical to that found in the traditional treatment, originally
due to Kaplan and Nelson\cite{kapnel}, in which the meson
field {\it is not} to
be identified with the divergence of the axial vector current.

Below, we discuss how the above ideas apply
to the analysis of S-wave {\it pion} condensation.
We thus sidestep the issue of whether the usual assumptions
of smoothness of amplitudes used in conjunction with PCAC,
are at all good in the $\rm{SU(3)}$
sector, which they  evidently are not,
given the relatively large size of the kinematical region over which
the smoothness assumptions are to assumed to hold.
However, the conclusions reached are directly applicable to
the discussion of S-wave kaon condensation.
We consider tree level lagrangians throughout,
working to ${\cal O}(Q^2)$.
We illustrate our results for
homogeneous, isotropic, isospin symmetric
and spin-unpolarized nuclear matter,
and evaluate nucleon operators
in the mean field approximation, such that the corresponding
results hold modulo nuclear correlation corrections.
The Letter is organized as follows.  In Section 2, we review the
kinematics
of pion-nucleon scattering and amplitudes obtained from PCAC.
We outline two approaches to chiral lagrangians, one originating
in the work of Manohar and Georgi\cite{mg}, and the other
in that of Gasser and Leutwyler\cite{gl}.
We discuss how the derived pion-nucleon scattering amplitudes
relate to the PCAC amplitudes.
In Section 3, we show that the effective pion
mass obtained in both formalisms give identical results.
The section concludes with a discussion of the Gell-Mann-Oakes-Renner
relation\cite{gmor} and a summary of results.

%%%%%%%%%%%%%%%%%%%%%%%%%%%%%%%%%%%%%%%%%%%%%%%%%%%%%%%%%%%%%%%%
%
%	Section 2
%
%%%%%%%%%%%%%%%%%%%%%%%%%%%%%%%%%%%%%%%%%%%%%%%%%%%%%%%%%%%%%%%%

\section{Off-mass-shell Pion-Nucleon Scattering \newline
Amplitudes and Chiral Lagrangians}

\indent
Low-energy pion-nucleon scattering was widely discussed in the
60s, when ideas such as current algebra and PCAC were developed.
A very convenient choice of kinematical variables in  the
description of pion-nucleon
scattering is the set $\nu$, $\nu_B$, $q^2$,
and ${q'}^2$, where
\beq
 \nu = \frac{s-u}{4 m_N}
 =\frac{(p+p')\cdot(q+q')}{4 m_N} \,\,\,{\rm and}\,\,\,
 \nu_B = \frac{ t^2-q^2-{q'}^2 }{ 4 m_N } = - \frac{q\cdot q'}{2 m_N}
 \,\,\,,
\eeq
\ni
in terms of the incoming (~outgoing~) pion four-momentum,
$q$ (~$q'$~) and nucleon
momentum $p$ (~$p'$~), and the Mandelstam variables
$s=(p+q)^2$, $t=(q-q')^2$,
$u=(p'-q)^2$. In what follows, the nucleon, of mass $m_N$ is taken
on mass-shell,
while the pion, of mass $m_\pi$, may be off mass-shell.
We consider the isospin even
pion nucleon scattering amplitude with
the pseudovector pole term subtracted, ${\bar D}^+$.
In the standard decomposition this amplitude reads\cite{hohler}
\be
\bar{D}^+( \nu, \nu_B, q^2, q'^2 ) &=& A^+( \nu, \nu_B, q^2, q'^2 )
+ \nu B^+( \nu, \nu_B, q^2, q'^2 ) \nno\\
&& \qquad\qquad - \frac{ \nu_B^2 }{ \nu_B^2 - \nu^2 }
\frac{g^2}{m}  K( q^2 ) K( q'^2 )      \,\,,
\ee
where g and $K(q^2)$ are, respectively, the $\pi NN$ coupling
constant and vertex form factor.
The amplitude ${\bar D}^+$ is closely related to the pion-nucleon
sigma commutator,
\be
 \sigma_{\pi N}(t) &=& \third \sum_{a=1}^3
 \langle P(p') | [ Q_5^a, [Q_5^a, {\cal H}_{\rm{SB}}] ]
  | P(p) \rangle \nonumber \\
 &=& \frac{m_u\mbox{+}m_d}2 \langle P(p') |
 \bar{u}u\mbox{+}\bar{d}d | P(p) \rangle ,
\ee
where $|P(p)\rangle$ is the proton state of
four-momentum $p$, normalized to 1,
$Q_5^a$ is the chiral charge of flavour $a$, and
${\cal H}_{\rm{SB}}=(m_u+m_d)(\bar{u}u+\bar{d}d)/2+m_s\bar{s}s$
is the explicit-chiral-symmetry-breaking Hamiltonian density,
with $m_u$, $m_d$ and
$m_s$ the masses of the up, down and strange quarks, respectively.
There are many choices for the pion field, and the one conventionally
used is the PCAC choice,
\beq
\pi^a = \frac{1}{f_\pi m_\pi^2} \del^\mu A^a_\mu
\,\,\,,
\label{pcac}
\eeq
where $f_\pi=93$ MeV is the pion decay constant and $A_\mu$ is the
axial vector current. Under the assumption of a smoothly varying
amplitude $\bar{D}^+$, one obtains from Eq.(\ref{pcac}) the simple
expressions at the following special
kinematical points\cite{chengdash,adldash}
\be
{\rm Weinberg\,\,point } &
\bar{D}^+(0,0,0,0) &= - \frac{1}{f_\pi^2} \sigma_{\pi N}(0)
\,\,\,,
\label{wb}
\\
{\rm Adler\,\,point} &
\bar{D}^+(0,0,m_\pi^2,0) &= \bar{D}^+(0,0,0,m_\pi^2) = 0
\,\,\,,
\label{adler}
\\
{\rm Cheng-Dashen\,\,point} &
\bar{D}^+(0,0,m_\pi^2,m_\pi^2)
&= + \frac{1}{f_\pi^2} \sigma_{\pi N}(0)
   +{\cal O}(m_\pi^3)       \,\,\,.
\label{cd}
\ee
The amplitude in the
Weinberg point is repulsive and equal in magnitude to the
attractive amplitude in the Cheng-Dashen point.
The Cheng-Dashen point is distinguished from the others
in that both pions are on mass-shell.
The sigma-commutator in Eq.(\ref{cd})
is evaluated at $t=0$ instead of
$t=2 m_\pi^2$ (the relevant value for the Cheng-Dashen point) as the
difference is of order $m_\pi^3$, and therefore outside the
scope of the present discussion. This will be detailed below.

Chiral perturbation theory was developed as a means to
obtain Green functions of QCD via an expansion in powers of
of the light quark masses and of the external
momenta, assumed to be small in relation to the
scale of chiral symmetry breaking $\Lambda \approx 1$GeV.
Here, we shall discuss the predictions
of such theories to second order (~next-to-leading order~)
in the characteristic momenta.
To illustrate our arguments regarding off-mass-shell amplitudes
and the effective meson mass in nuclear matter,
we examine two formulations
of chiral perturbation theory with baryons.
The first of these originates
in the formalism of Manohar and Georgi\cite{mg},
and was employed by Kaplan and Nelson\cite{kapnel} in proposing that
kaons could form a condensate.
The second formulation originates
in that of Gasser and Leutwyler\cite{gl},
which was extended to include baryons by
Gasser, Sainio and \u{S}varc\cite{gss}.
A different approach to the baryon sector,
using static baryon fields, was formulated
by Jenkins and Manohar\cite{jenman}.

We briefly review the chiral effective
lagrangian, ${\cal L}_{\rm{MG}}$,
originating in the formalism of
Manohar and Georgi\cite{mg}, including
the next-to-leading order terms in
chiral perturbation theory\cite{kapnel,jenman,bkrt}.
The lagrangian ${\cal L}_{\rm{MG}}$
is constructed from a
3 $\times$ 3 matrix, $\Sigma$, involving the pseudoscalar meson octet,
and the spin-$\frac12$ baryon octet matrix, $B$.
Under \suthree, the field transformations are
$\Sigma \rightarrow L \Sigma R^\dagger$,
and $B \rightarrow h B h^\dagger$, where $h$ is defined by the
transformation law $\xi \rightarrow L \xi h^\dagger=h \xi R^\dagger$,
and where the field $\xi$ is given by $\xi^2=\Sigma$.
For the purpose of constructing the lagrangian, the current
quark mass matrix, $M={\rm diag}(m_u,m_d,m_s)$,
is assigned the \suthree transformation of $\Sigma$.
The chiral expansion, in powers
of $\del_\mu \Sigma/\Lambda$ and $M/\Lambda$,
where $\Lambda$ is the chiral symmetry breaking scale,
$\Lambda \approx 1$ GeV, is constructed so that individual terms in
the expansion are \suthree invariant.
Baryons are treated as static, in the formalism first applied
to the chiral expansion by Jenkins and Manohar\cite{jenman}.
With a reduction to \sutwo
and the S-wave channel, the ${\cal O}(Q^2)$
lagrangian reads\cite{blrt,bkm}
\be
{\cal L}_{\rm{MG}} &=&
i \bar{N}\left( v \cdot \del \right)N
- \sigma \bar{N}N
+ \frac12 \left(\del_\mu {\bf \pi}\right)^2
- \frac12 m_\pi^2 {\bf \pi}^2 \nonumber\\
&+& \frac{1}{f_\pi^2} \left(
\frac12 \sigma {\bf \pi}^2 +
c_2 (v\cdot\del{\bf \pi})^2 + c_3 (\del_\mu{\bf \pi})^2 \right)
\bar{N}N
+ \cdots \,\,\,\,.
\label{lmg}
\ee
The quantity
$v_\mu$ is the four-velocity of the nucleon,  and
and reduces to $v_\mu=(1,0,0,0)$
in the  rest frame of the nucleon.
The pion mass is
$m_\pi=139$ MeV,
and $f_\pi=93$ MeV is the pion decay constant.
The values $m_\pi^2$ and $f_\pi^2$ include ${\cal O}(Q^2)$
loop corrections to the corresponding quantities at tree level,
and are used here for notational convenience, since we
neglect corrections to the lagrangian of ${\cal O}(Q^3)$ and higher.
The constants $\sigma$, $c_2$, and $c_3$ are linear in the
quark masses and therefore of ${\cal O}(Q^2)$.
The constant
$\sigma$ is to be identified with the sigma term, $\sigma(t=0)$,
which also serves to increase
the nucleon mass over that in the $\rm{SU(2)}$
chiral limit, $m_N=m_0+\sigma$,
where $m_0 \simeq 890$ MeV, using $\sigma= 45$MeV\cite{gls}.
Thus the sign of sigma is fixed to be positive.
We do not write down the Weinberg ( vector ) term
explicitly in the lagrangian above, as it does not enter
into the isospin even scattering length, or in the pion
self-energy in symmetric nuclear matter, to be considered
in Section 3. From
the lagrangian (\ref{lmg}), we find the isospin even scattering
length, $a^+_{\pi N}$,
\beq
 a^+_{\pi N} =
 \frac{1}{4\pi f_\pi^2 (1+m_\pi/m_N)}
     \left(2(c_2+c_3)m_\pi^2+\sigma\right)
 + \om3 \,\,\,.
\eeq
Empirically,  $a^+_{\pi N} = -0.0083 m_\pi^{-1}$ \cite{koch},
corresponding to a repulsive interaction.
Using $\sigma = 45$ MeV, we find $(c_2+c_3)m_\pi^2 = -26$~MeV.
Improved values for  the constants can be found by including
loop corrections \cite{bkm}: There are terms of ${\cal O}(m_\pi^3)$
resulting from finite loop corrections, whereas
infinite loop corrections and therefore also
the corresponding counterterms  first appear at ${\cal O}(m_\pi^4)$
and ${\cal O}(m_\pi^4 \ln m_\pi)$.
This is the reason why in nuclear matter
all  quantities
of ${\cal O}(Q^2)$ get their first correction already at
${\cal O}(Q^3)$ and not at  ${\cal O}(Q^4)$ as their free-space
analogs.
The lagrangian(\ref{lmg}) gives the amplitude
\beq
 \bar{D}^+ = + \frac{1}{f_\pi^2} \sigma + {\cal O}(m_\pi^3)
 \label{ampmg}
\eeq
\ni in all three kinematical points discussed
above, in apparent contradiction
with the PCAC result.
This statement has been given various interpretations,
for example that chiral perturbation
theory is incorrect off mass-shell,
or in contradiction with well established extrapolations such as
PCAC\cite{dee,sc1,sc2}.

With regard to the disagreement between Eq.(\ref{ampmg}) and
Eqs.(\ref{wb},\ref{adler}), it must be kept in mind that
the amplitudes (\ref{wb},\ref{adler}) are not fundamental, in the
following sense.
Any operator may be adopted to describe the pion field, provided
that it is correctly normalized, and has a nonvanishing
matrix element between the pion and the vacuum.
Amplitudes obtained off-pion-mass shell are subject to the
choice adopted for the pion operator and are not unique.
(~However, they are
not entirely arbitrary, being subject to constraints
such as unitarity and causality.~)
This does not pose a problem,
since the off-pion-mass-shell amplitudes
are {\em not\/} accessible by experiment, and as shown
by  Coleman, Wess and Zumino\cite{ccwz}, S-matrix elements,
and therefore measured scattering amplitudes, are independent
of the choice adopted for the pion field.
The result (\ref{ampmg}) is therefore simply to be understood
in the way that the pion field $\bf \pi$ of ${\cal L}_{\rm{MG}}$
is not the PCAC one.

One should  furthermore note that the
amplitudes (\ref{ampmg}) and (\ref{cd})
agree, to ${\cal O}(Q^2)$, as required,
since the Cheng-Dashen point has the
virtue of being on pion-mass-shell.
Although the Cheng-Dashen point is unphysical in the sense that
it is outside the kinematical region available to pion-nucleon
scattering experiments, the scattering amplitude in the
Cheng-Dashen point is unique and related to measured amplitudes
via dispersion relations. These involve analytic continuations
of (on-pion-mass-shell) amplitudes to unphysical regions --  unique
by Cauchy's theorem,
provided the various poles are properly subtracted.  A commonly
used approach is to perform double dispersion relations
in the variables $(\nu,t)$\cite{hohler,gss},
the degrees of freedom being reduced by two, on pion-mass-shell.
A variation in pion-mass can, of course, not be carried out
in any experiment; so the Weinberg and Adler points are unphysical
in a very different sense than the Cheng-Dashen point.

To conclude this section, we review
the evaluation of off-shell scattering
amplitudes in the functional integral
formulation of chiral perturbation theory developed by Gasser
and Leutwyler\cite{gl}, which was extended to include nucleons by
Gasser, Sainio and \u{S}varc\cite{gss}.
In this approach, a generating functional for Green functions,
$Z_{\rm{eff}}$,
and the corresponding lagrangian,
${\cal L}_{\rm{GSS}}$, are developed as follows.
External sources, $s,p,v_\mu$,{\rm and} $a_\mu$,
are coupled to the quark fields of the QCD action and assigned
chiral \sutwo transformations such that the source-extended  action
is {\it locally}
chiral invariant.  The fact that the sources are coupled in this
way ensures that chiral QCD-Ward identities are satisfied.
The generating functional for the low-energy effective theory  --
transcribed to the hadronic level --
now depends on the {\em same\/} external sources,
where the corresponding hadronic action
is formulated in terms of a 2 $\times$ 2
dummy field $U$, involving mesons.
It is a dummy field as it is integrated over
in the generating functional formulation.
Similarly, one may introduce nucleons, $N$($\bar N$) coupled to
corresponding external sources, $\bar\eta$ ($\eta$),
which transform non-linearly under
chiral transformations in order to ensure that the coupling terms are
(locally) chiral invariant.
One identifies the Ward identities
of the low-energy effective theory with those of QCD.
Note that in this formalism,
the sources $s$, $p$, $v_\mu$, and $a_\mu$
are coupled from the start
to quark bilinears and are therefore directly associated with
\sutwo currents in quark variables.
The PCAC prescription thus emerges naturally as the sources are
transcribed to the hadronic level.
The relevant generating functional is
\beq
e^{i  Z_{\rm{eff}}[s,p,v_\mu,a_\mu,\eta,{\bar \eta}]}
= {\cal N}
\int d U \, dN \, d{\bar N}
e^{i  \int\! d^4x\, ( {\cal L}_{\pi \pi} + {\cal L}_{\pi N} +
{\bar \eta}N + {\bar N}\eta ) }
\,\,\,,
\label{zeff}
\eeq
\ni where $\cal N$ is an overall normalization, $U,N,{\bar N}$
are dummy fields, and the lagrangians $ {\cal L}_{\pi \pi} $
and $   {\cal L}_{\pi N} $ are dependent on
the sources $s$, $p$, $v_\mu$ and $a_\mu$.
In what follows, we shall use the scalar source $s$ to generate the
quark mass matrix, $s={\cal M}={\rm diag}(m_u,m_d)$,
and furthermore retain only the
source for asymptotic pions, $p$,
but set $v_\mu=a_\mu=0$ (~unless otherwise specified~).
The sources $\bar \eta$ and $\eta$ generate
one-nucleon in- and out-states.  As before, the nucleons are
treated in the static fermion formalism\cite{jenman},
in which nucleon loops play no role
( see also Appendix A of Ref.\cite{bkkm} ), and the nucleon
determinant may therefore taken to be unity.

The lagrangian entering in the generating functional (\ref{zeff}),
${\cal L}_{\rm{GSS}}={\cal L}_{\pi \pi} + {\cal L}_{\pi N}$,
is to leading order given by the nucleon kinetic energy term
and to subleading order, ${\cal O}(Q^2)$, by\cite{gss,bkm}
\beq
{\cal L}_{\pi \pi}^{(2)} =
\frac{f_\pi^2}{4} {\rm Tr} \del_\mu U \del^\mu U^\dagger
+ \frac{f_\pi^2}{4} {\rm Tr} ( U^\dagger \chi + \chi^\dagger U )
\,\,\,,
\label{lpp2}
\eeq
and,
\beq
{\cal L}_{\pi N}^{(2)} =
- \frac{\sigma}{4m_\pi^2} {\bar N}N {\rm Tr}
( U^\dagger \chi + \chi^\dagger U )
+ c_2 {\bar N}(v \cdot u)^2 N + c_3 {\bar N} (u \cdot u) N
\,\,\,,
\label{lpn2}
\eeq
where $u_\mu=i u^\dagger \del_\mu U u^\dagger$,
$U=u^2=\exp(i \tau^a \pi^a/f_\pi)$,
$\chi=2 B (s+i p)$, and $v_\mu$ is the four-velocity of the
nucleon.
Expanding $U$ to second order in $\pi^a$, we find
\beq
{\cal L}_{\rm{GSS}} = {\cal L}_{\rm{MG}} + j^a \pi^a
\left( 1 - \frac{\sigma {\bar N}N}{f_\pi^2 m_\pi^2}\right )
\,\,\,.
\label{lgl}
\eeq
The pseudoscalar source is $j^a=2 B f_\pi p^a$,
in terms of the original
source $p^a$ ($p=p^a \tau^a$),
and of the ``quark condensate''
$-2 f_\pi^2 B=-2 f_\pi^2 m_\pi^2/(m_u+m_d)$,
as  follows from the quadratic expansion.
Since Green functions are obtained by taking functional derivatives
of the generating functional with respect to the source $j^a$,
the nontrivial coupling of the source
to the pion field in Eq.(\ref{lgl})  plays an important
role.

Solving the equation of motion for the canonical field in terms
of the sources, one easily obtains the Ward identity\cite{gss}
\be
  i \int d^4 x \langle N(p') | T P^a(x) P^b(y) | N(p)
  \rangle e^{ iq'x - iqy }
 \nno \\
=
\frac{ g_\pi }{ m_\pi^2 - q^2 } \frac{ g_\pi }{ m_\pi^2 - q'^2 }
T_{\pi N}^{a b}( p'q';pq )  \,\,\,,
\label{ward}
\ee
where
\beq
T_{\pi N}^{a b} = T_{pv}^{a b} +
\left\{
\frac{ q^2 + q'^2 - m_\pi^2 }{ f_\pi ^2 m_\pi^2 } \sigma
+ 2 \frac{c_2 \omega \omega' + c_3 q \cdot q'}{f_\pi^2}
\right\} \delta^{a b}
\,\,\,,
\label{tamp}
\eeq
and
\beq
 g_\pi \delta^{ab} =
 \langle 0 | P^a | \pi^b \rangle = 2 B f_\pi \delta^{ab}
\,\,\,,
\label{gpi}
\eeq
with $P^a={\bar q}i \gamma_5 \tau^a q$ in terms of quark fields.
Equivalently, the amplitude (\ref{tamp}) follows from the
connected $\pi \pi {\bar N} N$ Green function
with amputated nucleon legs,
$A_{\pi N} (q^2,q'^2)$.
The function $A_{\pi N} (q^2,q'^2)$ is given by
the Feynman diagrams of Fig.~1.
Fig.~1a) represents the three contributions  having
two external pion legs.
The relevant vertices follow from
the interactions in  lagrangian (\ref{lgl}) quadratic
in pion fields, proportional to $\sigma$, $c_1$ and $c_2$.
Figs.~1b) and 1c) represent the contributions to
$A_{\pi N}$ containing only one external pion leg.
These terms have weight $-\sigma/(f_\pi^2 m_\pi^2)$ and are
due to the coupling of the source $j^a$ to the nucleon
fields in Eq.(\ref{lgl}).
The sum of all these contributions  is
\be
A_{\pi N}(q^2,q'^2) =
\frac{i }{q^2-m_\pi^2}
\left(
\frac{i\, \sigma}{f_\pi^2} +
\frac{2i }{f_\pi^2}
\left( \omega \omega' c_2 + q \cdot q' c_3 \right)
\right)
\frac{i }{q'^2-m_\pi^2}
\nonumber \\
\,\,\,\,\,\,\,\,\,\,\,
+
\left( - \frac{\sigma}{f_\pi^2 m_\pi^2} \right)
\frac{i }{q'^2-m_\pi^2}
+
\frac{i }{q^2-m_\pi^2}
\left( - \frac{\sigma}{f_\pi^2 m_\pi^2} \right)
\,\,\,.
\label{vacdiag}
\ee
Amputating external pion legs by multiplying with $i (q^2-m_\pi^2)
(q'^2-m_\pi^2)$, we obtain the second term on the r.h.s. of amplitude
(\ref{tamp}).
It agrees with the amplitudes (\ref{wb},
\ref{adler}, \ref{cd}), demonstrating
explicitly that there is no inconsistency
between  chiral
perturbation theory and  PCAC.

One may be tempted to raise the objection
that the amplitude (\ref{ampmg})
is inconsistent with the Ward identity (\ref{ward}),
which is presumably
a direct consequence of the underlying theory, QCD.
However, it must be kept in mind that the Eq.(\ref{ward}) is
a statement about isovector pseudoscalar currents,
which are convention-dependent
in contrast to vector or axialvector currents
which have a physical realization, the
electro-weak currents.
In the language of PCAC, one arrives
at the amplitudes  (\ref{wb},
\ref{adler}, \ref{cd}) by identifying the
pion interpolating field with the
divergence of the axial-current, Eq.(\ref{pcac}).
In the formulation
of Manohar and Georgi\cite{mg} this identification is not made,
as is evidenced by Eq.(\ref{ampmg}).  This difference is
of no consequence for
the evaluation of S-matrix elements\cite{ccwz},
nor in the evaluation
the effective mass (~Section 3~).

We close this section with a few additional comments on the relation
between the two formulations outlined above.
First, we remark that although the second of the
approaches described above is based on functional integrals,
while the first is not, this is purely a notational distinction.
We could just as well have derived the amplitude (\ref{ampmg}), in
a functional integral formalism, using the action
$S= \int d^4 x ( {\cal L}_{\rm{MG}} + \pi^a J^a )$, by taking
derivatives of the corresponding generating functional
with respect to $J^a$. The two approaches differ in the
structure and coupling of their external sources.
Second, there are complications arising at the one-loop level.
At the one-loop level, in computing four-point and higher
Green functions of the basic
fields in models such as the non-linear sigma model,
one encounters divergences off-mass-shell that cannot be removed
-- in a standard way -- by chiral
invariant counterterms. These divergences vanish on-mass-shell
and may, more generally, be removed by a suitable redefinition of
fields which have to be reexpressed by a non-linear combination of
the original fields {\em including\/}
space-time derivatives \cite{redef}.
Introducing external fields via covariant derivatives in a chiral
invariant manner in the generating functional framework,
one obtains n-point Green functions involving currents,
and chiral invariance is directly  obtained to a given
order in the loop expansion\cite{honerkamp,gl}.
In other words, this scheme generates by construction  counter
terms which are non-linear in the external sources
(or contain space-time derivatives thereof)
and which one otherwise would have
to provide by hand in order to redefine
the original fields in the first scheme.
The point is once more: in the second scheme
one has a relationship (dictated by the local chiral symmetry)
between the usual terms in the lagrangian and
the structures of the sources.
By a redefinition (wave function renormalization) of the pion field
one can make the source structures trivial, however, at
the expense of a rather complicated non-linear
field redefinition for the pion field
which involves derivatives as the pseudo scalar source
involved derivatives.
There  is of course -- also in the second scheme --
an arbitrariness in
introducing the sources on the QCD-level such that
chiral symmetry is locally preserved. But once this has been fixed,
there is no arbitrariness any longer
in the transcription down from the QCD level to the hadronic one.
In this line of thought, the first scheme
(especially at higher loop order)
would
correspond to a non-standard and rather complicated coupling
of the external sources to the current quarks at the QCD level.

In the present work, we concentrate on tree level calculations,
Green
functions or correlators  with at most  two external pion legs, and
the above complications do not arise.  Further work is needed
to clarify the situation at one-loop order,  but a recent
calculation\cite{mannque} indicates that the two approaches
discussed here also give the same effective meson mass to one-loop
order, as the pion field redefinitions only contribute
at higher order than
${\cal O}(\pi^2)$.

%%%%%%%%%%%%%%%%%%%%%%%%%%%%%%%%%%%%%%%%%%%%%%%%%%%%%%%%%%%%%%%%%%%%%
%
%	Section 3
%
%%%%%%%%%%%%%%%%%%%%%%%%%%%%%%%%%%%%%%%%%%%%%%%%%%%%%%%%%%%%%%%%%%%%%

\section{The Effective Meson Mass in Nuclear Matter}

\indent
Given the problems encountered in extending
chiral perturbation theory from the meson to the baryon sector,
it is not surprising that a rigorous formulation
of the expansion in nuclear matter has not yet been found.
The presence of an additional scale,
the Fermi momentum of nucleons, the breaking of Lorentz invariance,
and nuclear correlations, add
new levels of complexity to the formulation of a chiral expansion.
As a first step, one simply uses a free space chiral expansion,
such as those outlined above
and evaluates nucleon operators at the mean field level, and
consequently works with the lagrangian to linear order in density.
One may again raise the question of the role of the pion
interpolating field in this context.  In what follows, we will argue
that the basic idea, established rigourously
in the case of free space scattering, that physically relevant
observables are independent of the choice of field variables,
also holds in nuclear matter. In this case, the relevant
observable is the position of the pole of the pion propagator
in symmetric nuclear matter\cite{polerefs,tpl}.
The pole position is often referred to as the effective mass,
which we shall also do here.
In the absence of a general proof that the
poles of the propagator are independent of the
choice of field variables, we demonstrate that the two lagrangians
discussed above, and most often employed in chiral perturbation
theory, do indeed give the same value for the effective pion in mass
in nuclear matter.

As alluded to above, in the nucleon mean field approximation
we set $\langle{\bar N}N\rangle=\rho$ (~we also approximate
the vector density by the scalar density.~) From the
Manohar-Georgi lagrangian (\ref{lmg}), we thereby obtain the
inverse propagator for charged pions
\be
D^{-1}(\omega, {\bf k},\rho) &=&
\omega^2 - {\bf k}^2 - m_\pi^2 - \Pi(\omega,{\bf k},\rho) \nonumber\\
\Pi(\omega,{\bf k},\rho) &=& - \frac{\rho}{f_\pi^2}\,
\left( 2(c_2+c_3)\omega^2-2c_3{\bf k}^2+\sigma\right) +\om3
\,\,\,\,.
\label{mgprop}
\ee
Evaluating the poles
of the propagator we find the effective pion mass
${m_\pi^\ast}^2(\rho):=\mbox{$\omega^2({\bf k}=0;\rho)$}$ in
symmetric nuclear matter to be
\be
{m_\pi^\ast}^2(\rho)
&=& m_\pi^2 \frac{1-\frac{1}{f_\pi^2 m_\pi^2}\,\sigma\,\rho}
                       {1+\frac{2}{f_\pi^2}\,(c_2+c_3)\,\rho}  +\om3
\label{mpifull} \\
       &=& m_\pi^2\left( 1 - \frac{\rho}{f_\pi^2 m_\pi^2}(
2(c_2+c_3)m_\pi^2
+\sigma) \right )
           + {\cal O}(m_\pi^3;\rho^2)
\,\,\,,
\label{mpilin}
\ee
\ni where in Eq.~(\ref{mpilin}) we explicitly show
the prediction to linear order in
density.
Eq.~(\ref{mpilin}) gives
$m_\pi^\ast(\rho) = m_\pi - 2 \pi a^+_{\pi N} \rho / m_R$, to
linear order in density, where
$m_R$ is the reduced mass of the pion-nucleon system.
The behaviour of the effective mass to linear order in
density is as expected from the lowest order optical
potential,  without reference to chiral perturbation theory.
The effective mass receives additional contributions
over those given by Eq.(\ref{mpifull}) at higher than
linear order in density,
from factors such as those mentioned at the beginning of this
section.
In the equations above, and in those
which follow, we exhibit explicitly only the density
dependence following from the chiral lagrangian,
to ${\cal O}(Q^2)$,  and in the mean-field approximation.

Similarly, we consider the lagrangian ${\cal L}_{\rm{GSS}}$,
Eq.(\ref{lgl}),
in the mean field
approximation.  Note that from (\ref{zeff}) it is immediately
apparent that
since $U$, and therefore $\pi$, is a dummy variable, any
observable must be independent of this field.
In what follows, we obtain the effective pion mass in symmetric
nuclear matter
by two (~equivalent~)
methods, the first uses the effective
action $\Gamma$, (~see e.g.\ Ref.\cite{pokorski}~),
and the second, Feynman diagrams.
In obtaining the effective action,
the so-called classical pion field $\phi_\pi$ is
defined with the help of the
the generating functional in the nucleon mean field
approximation, $Z[j,\rho]:=Z_{\rm{eff}}[{\cal M},j,0,0,0,0]$,
as follows:
\beq
\frac{ \delta Z [ j, \rho ] }{ \delta j^a} = \phi_\pi^a
\label{phidef}
\,\,\,.
\eeq
At tree level, we have
\beq
Z[j,\rho] = S[\pi] + \int d^4 x j^a(x) \pi^a(x)
\left( 1 - \frac{\sigma \rho}{ f_\pi^2 m_\pi^2 } \right)
\,\,\,,
\eeq
where the action $S$ is given by the lagrangian
${\cal L}_{\rm{GSS}}$
in the nucleon mean field approximation, without the source $j^a$,
namely,
$S=\int d^4 x {\cal L}^{(2)}(\rho)$ with
\beq
{\cal L}^{(2)}(\rho)  =
	  \frac{f_\pi^2}{4}\left(g^{\mu\nu}
       \mbox{+}\frac{D^{\mu\nu}\rho}{f_\pi^2} \right )\! {\rm Tr}
          (\del_\mu U \del_\nu U^\dagger )
       + \frac{f_\pi^2}{4} \left ( 1\! -\!
 \frac{\sigma\rho}{f_\pi^2m_\pi^2}\right )\!
          {\rm Tr}(U^\dagger \chi\mbox{+}\chi^\dagger U),
\label{lrho}
\eeq
and $D^{\mu \nu} \equiv 2 c_2 v^\mu v^\nu + 2 c_3 g^{\mu \nu}$,
as follows
from Eqs.(\ref{lpp2}) and (\ref{lpn2}). From Eq.(\ref{phidef}),
we obtain
\beq
\phi_\pi^a
= \left( 1 - \frac{\sigma \rho}{ f_\pi^2 m_\pi^2 } \right)
\pi^a
\,\,\,.
\eeq
The tree-level effective action,
$\Gamma[\phi_\pi,\rho]=Z[\phi_\pi,\rho]-
\int d^4x j \phi_\pi =
S[(1-\frac{\sigma\rho}{f_\pi^2 m_\pi^2})^{-1}\phi_\pi]$,
therefore reads
\be
&&\Gamma[\phi_\pi,\rho] =
\nonumber \\
& &  \half \int d^4x\,
\left( 1 - \frac{\sigma \rho}{ f_\pi^2 m_\pi^2 } \right)^{\!-2}\!
\left(
\del_\mu \phi_\pi \del_\nu \phi_\pi
\left( g^{\mu \nu}\mbox{+}\frac{D^{\mu \nu} \rho }{f_\pi^2} \right)
- m_\pi^2
\left( 1\!
-\! \frac{\sigma \rho}{ f_\pi^2 m_\pi^2 } \right)\! \phi_\pi^2
\right)\,\,, \nonumber \\
\ee
\ni from which we obtain the in-medium charged pion propagator
\beq
D(q,\rho) =
\frac{ i \left( 1
- \frac{\sigma \rho}{ f_\pi^2 m_\pi^2 } \right)^2 }
{ q^2 - m_\pi^2 + \frac{\rho}{f_\pi^2}
( \sigma + 2 c_2 (v \cdot q)^2 + 2 c_3 q^2 ) } +\om3 \ .
\label{gssprop}
\eeq
As with the example of  pion-nucleon scattering
amplitudes considered earlier, it is instructive to rederive these
results with the help of  Feynman diagrams.  From
the lagrangian (\ref{lgl})
in the mean field approximation,
we obtain two vertices,  $\odot$
and $\otimes$, connecting to
two bare pion lines and one bare pion line, respectively.
The vertices are
\beq
\odot =  i \frac{\rho}{f_\pi^2}
 ( \sigma + 2 c_2 (v \cdot q)^2 + 2 c_3 q^2 )
\,\,\,\,\,\,\,
{\rm and}
\,\,\,\,\,\,\,
\otimes = -\frac{\sigma \rho}{ f_\pi^2 m_\pi^2}
\,\,\,.
\eeq
The in-medium pion propagator follows from summing
all possible contributions dressing the bare pion
propagator, as shown in the first
line of Fig.~2a).
The contributions involving $\odot$ yield a geometric series
whose sum
is denoted by the double dashed line in Fig.~2b).
This yields the four diagrams of the second line of Fig.~2a), which
factorize to yield the single expression (\ref{gssprop}).

The poles of the pion propagator in symmetric nuclear matter,
(\ref{gssprop}),
obtained from the lagrangian ${\cal L}_{\rm{GSS}}$, are
identical to those
of the propagator (\ref{mgprop}),  following from the lagrangian
${\cal L}_{\rm{MG}}$.
In both cases, the effective pion mass is given by
Eq.(\ref{mpifull}).   The fact that the lagrangian
${\cal L}_{\rm{GSS}}$
emodies the PCAC choice of pion interpolating field  appears
in the residue of the pion propagator,  not in the effective
pion mass.
We therefore conclude that  there is no discrepancy in the
effective mass obtained from theories using different
interpolating pion  fields\cite{sc1,sc2}.
It is furthermore to be noted that,  within the approximations that
we are working with and have stated  above, the effective mass
predicted by both models is identical {\it to all orders in nuclear
density}. From the the tree-level calculation of the  effective pion
mass, one cannot conclude that the two
approaches are intrinsically different,  to higher than linear
order in density\cite{sc2}.

%%%%%%%%%%%%%%%%%%%%%%%%%%%%%%%%%%%%%%%%%%%%%%%%%%%%%%%%%%%%%%%%%%%%%
%
%	GMOR
%
%%%%%%%%%%%%%%%%%%%%%%%%%%%%%%%%%%%%%%%%%%%%%%%%%%%%%%%%%%%%%%%%%%%%%

To conclude this section, we discuss the relation of the
effective pion mass in symmetric nuclear matter,
to the Gell-Mann-Oakes-Renner
(~GMOR~) relation\cite{gmor}.
The Gell-Mann-Oakes-Renner (~GMOR~)
relation reads
\beq
f_\pi^2 m_\pi^2 = -\frac{m_u+m_d}{2}
\langle 0 | \bar{u}u+\bar{d}d | 0 \rangle
+ {\cal O}(m_\pi^4)
\label{gmor}
\eeq
\ni
The GMOR relation is
derived from the effective second-order lagrangian
${\cal L}^{(2)}_{\pi \pi}$, Eq.(\ref{lpp2}), by the identification
$Z_{\rm{QCD}}=Z_{\rm{eff}}$, which leads to
\beq
 \left. \frac{\delta Z_{\rm{QCD}} }{\delta {\cal M}(x)}
  \right|_{{\cal M}=0}
  = - \langle 0 | \bar u u + \bar d d | 0 \rangle = 2 f_\pi^2 B
 \label{findgmor}
\eeq
Solving for $B$ and using the relation  $B (m_u+m_d) = m_\pi^2$
one thus obtains  Eq.(\ref{gmor})\cite{gl}.
To study the GMOR relation with respect to the in-medium pion mass,
we use the lagrangian ${\cal L}_{\rm{GSS}}$
in the nucleon mean field approximation, Eq.(\ref{lrho}).
Since matter breaks Lorentz invariance (but still
keeps rotational invariance,
if it is isotropic),
it is convenient to separate
space and time components \cite{kirch} via
(~see also Ref.\cite{leutnonrel}~),
\beq
f_\pi^2 \left(g^{\mu\nu}
+ \frac{D^{\mu\nu}\rho}{f_\pi^2} \right )
= {f_t^\star}^2(\rho)g^{00} + {f_s^\star}^2(\rho) g^{ii}
\,\,\,,
\eeq
where the time-component is given by
\beq
 {f_t^\star}^2 (\rho) = f_\pi^2
 \left(1 + \frac{ D^{00} \rho}{f_\pi^2}\right )
  + {\cal O}(m_\pi)  \ .
\label{ftrho}
\eeq
Starting from Eq.(\ref{lrho})
and using the same method as to derive Eq.(\ref{findgmor}),
we obtain the density dependent quark condensate
\beq
\langle \bar u u + \bar d d \rangle_\rho =
\langle 0 | \bar u u + \bar d d | 0 \rangle
\left( 1 - \frac{\sigma\rho}{f_\pi^2m_\pi^2}\right )
 +{\cal O}(m_\pi)
\,\,\,,
\label{qqrho}
\eeq
a result which is, in fact, model-independent\cite{birse}.
Eq.(\ref{qqrho}), when combined with the effective pion mass as
given in Eq.(\ref{mpifull}),
yields the in-medium GMOR relation in
the nucleon mean-field approximation:
\beq
{f_t^\star}^2(\rho)\, {m_\pi^\star}^2(\rho) = -\frac{m_u+m_d}{2}
\langle \bar{u}u+\bar{d}d \rangle_\rho
+ \om3 \,\,\,.
\label{gmorrho}
\eeq
(~Other discussions of the GMOR relation
at finite density are given in Ref.\cite{gmorrho}.~)
It is therefore only the time-component of the coupling constant,
$f_t^\star$
that enters in the GMOR relation at finite density.
As a function of density, $f_t^\star$ decreases, $m_\pi^\star$
increases, though
very slowly, and $-\langle \bar{u}u+\bar{d}d \rangle_\rho$ decreases.

It is worthwhile to check that $f_t^\star(\rho)$, as given
by Eq.(\ref{ftrho}),
agrees with the definition in terms of the axial current coupling
to the pion in matter,
\beq
\langle 0 | \bar q \gamma_0 \gamma_5 \frac12 \tau^a q
| \pi^b\rangle_\rho = i p_0 \delta^{ab} f_t^\star(\rho)
+ \om3
\,\,\,.
\label{ftgen}
\eeq
The expectation value may be evaluated from the axial vector two
point function, which reads, at zero nucleon density,
(see Ref.\cite{gl})
\be
\left. \frac{\delta^2  \Zeff}
{\delta a_\mu^a(-q)\,  \delta a_\nu^b(q)}
\right|_{a=v=p=0;s={\cal M}} &=&
 i   \int d x\,
e^{i  q (x-y)} \langle 0| T A_\mu^a (x) A_\nu^b(y) |
0\rangle \nno \\
 &=& \delta^{ab}\left \{ g_{\mu\nu} f_\pi ^2
+ \frac{q_\mu q_\nu f_\pi^2}
{\Mpi^2 -q^2}
\right \} + {\cal O}(q^2) \,\,\,. \label{aa}
\ee
Eq.(\ref{aa}), and other correlators, may be evaluated
at finite density, in the mean field approximation,
by reinstating the general sources
$s$, $v^\mu$ and $a^\mu$ in $Z_{\rm{eff}}$,
expanding the action to
second order in the pionic field, and integrating out the pionic
degrees of freedom.  The second order variation
with respect to the external sources (~e.g.\ $a_0^a$ for the case
in Eq.(\ref{aa})~) then gives the two-point function.
For example, one finds that -- up to ${\cal O}(m_\pi)$ corrections --
the time-component of the axialvector correlator
in matter has the same form as Eq.(\ref{aa}),
but with $f_\pi^2$ replaced by ${f_t^\ast}^2
(\rho)$, Eq.(\ref{ftrho}),
and $m_\pi$ replaced
by the effective mass (\ref{mpifull}).
Thus the desired equivalence
is established. This result is independent of the off-shell
extension of the
pion field, as the pseudoscalar sources, $p^a$, do not enter
in this calculation.

In an analogous manner one may arrive at other relations valid at
finite density.
Evaluation of the in-medium pseudoscalar correlator results in (see
Eq.(\ref{gpi})~)
\beq
 {g_\pi^\star}^2 (\rho) = (2 B f_\pi)^2
 \frac{    \left (1 -\frac{\sigma \rho}{f_\pi ^2 \Mpi^2} \right )^2 }
                    {1 +\frac{D^{00} \rho}{f_\pi^2}}
  + {\cal O}(m_\pi)  \ .
\label{gpirho}
\eeq
This result is
dependent on the off-shell extension of the pion field,
as the
calculation explicitly involves functional derivatives with respect
to the
pseudoscalar source, $p^a$, and holds
therefore only in the second scheme,
compatible with PCAC.
The axialvector-pseudoscalar correlator is scheme-dependent, as well.
Using Eqs.(\ref{qqrho}) and (\ref{mpifull})
one can then check that the finite-density
version of the PCAC relation (\ref{gpi})  holds in the second scheme:
\beq
  f_t^\star(\rho) {m_\pi^\star}^2(\rho)
 = \frac{m_u+m_d}{2} g_\pi^\star(\rho)
  + \om3 \ .
\eeq

%%%%%%%%%%%%%%%%%%%%%%%%%%%%%%%%%%%%%%%%%%%%%%%%%%%%%%%%%%%%%%%%%%%%

A  caveat is the rather rapid decrease of the in-medium quark
condensate, Eq.(\ref{qqrho}), (~which drops at nuclear matter density
already to about two thirds
of its vacuum value~), as a large value of the
quark condensate is a precondition on counting the current
quark mass matrix as ${\cal O}(Q^2)$. If the (in-medium)
quark condensate gets smaller and smaller,
it is not obvious that the (in-medium)
four-quark condensate can be safely neglected at ${\cal O}(Q^2)$.
It may become as important
as the two-quark condensate, such that the quark
mass matrix should rather be counted
as ${\cal O} (Q^1)$ (~see  Ref.\cite{ks} in a different context~).
Note that even in {\em symmetric\/} nuclear matter
all quantities of ${\cal O}(Q^2)$ get
their first corrections already at  ${\cal O}(Q^3)$ -- as the
mean-field calculation showed -- and not at
${\cal O}(Q^4)$ as their free-space analogs. One might speculate
that this fact is already a signal or precursor
of a change in the  relation between ${m_\pi^\star}^2 (\rho)$
and the current quark mass matrix $\cal M$  from that given by the
vacuum GMOR relation.
This phenomenon has  eventually
to take place with
increasing density,
as at chiral restoration (~or perhaps even earlier~)
the pion should lose its Goldstone character
and approximate the behavior of
a ``normal'' meson  with the corresponding relation  between
its (effective) mass and
the current quark masses:
$m_\pi^\ast \propto \hat{m}=(m_u+m_d)/2$ \cite{wt}.
Thus the in-medium GMOR relation (\ref{gmorrho}) is
likely to break down at
large nuclear densities.

To summarize the results of this Letter,
since the isospin even scattering length is
$a^+_{\pi N} = -0.01 m_\pi^2$,
the pion mass increases slightly to linear order in density
in symmetric nuclear matter\cite{dee}.
This result is valid at low density,  and is, of course, expected,
whether one uses a chiral effective lagrangian or not.
Tree-level chiral perturbation
theory in the nucleon mean field approximation
predicts that the effective mass continues to grow
with increasing density, as given by Eq.(\ref{mpifull}).
Eq.(\ref{mpifull}) was derived from two
formulations of chiral perturbation theory.
One of these\cite{gl,gss,bkm}
corresponds to adopting the PCAC choice of
pion interpolating field, while
the other\cite{mg} does not.
The fact that the two formulation give identical results
indicates that the effective mass in matter is generally
independent of the choice of pion field variables
adopted in the effective lagrangian.

To which extent these predictions are correct can only finally
be decided by better knowledge of higher density terms, both
in terms of nuclear correlations, and at the level of the lagrangian.
Confining the discussion to the chiral lagrangian, at tree-level,
and in the nucleon mean field approximation,
we have found that the two approaches outlined above give
identical effective pion mass to all orders in nuclear density.
This seriously questions the claim that differences
in the choice of pion field manifest themselves as different
predictions to second order in density and above\cite{sc2}.

A number of additional factors must be taken into consideration
in studying {\it kaon} condensation in neutron star matter, such
as the additional attraction mediated by the
exchange of the $\rho$ and $\omega$ vector mesons, i.e.\ the
Weinberg term.
Other pertinent issues include the role of resonances such as
the $\Lambda(1405)$, which governs low-energy $K^- p$
scattering
and the coupling to the $\Sigma \pi$
channel\cite{sc1,lambda},
the role of hyperons\cite{kapnel,hyperons},
the effect of ${\cal O}(Q^3)$ contributions
in chiral perturbation theory\cite{chlee} and
nuclear correlations\cite{chlee,ppt}.
These complications do not detract from our basic conclusion,
namely that the effective kaon mass, in neutron star matter,
is independent of the choice of kaonic field varmade in the effective
lagrangian.
Although kaon-nucleon scattering amplitudes, obtained
from a specific choice of kaon interpolating field, may be repulsive
off meson-mass-shell,
such as the PCAC amplitude in the Weinberg point,
this does not imply that that the effective kaon mass tends
to increase in matter, precluding kaon condensation.
In this respect, the effective kaon mass in matter, as obtained
originally by Kaplan and Nelson\cite{kapnel}, and in most studies
that have followed, has been correctly evaluated.

\vskip 5mm
\ni {\bf Acknowledgements}

\ni
We are grateful to F. Beck, M. Kirchbach,
M. Lutz,  and W. Weise for discussions and to
H. Bijnens, G. E. Brown, N. Kaiser, A. Manohar, U.-G. Mei{\ss}ner,
M. Rho and I. Zahed for
discussions and a critical
reading of a  previous version of the manuscript.
We thank
the ECT$^{*}$, Trento, Italy, for support
to attend its International Workshop on Chiral
Symmetry in Hadrons and Nuclei.
A.~W. would like to thank NORDITA and
the Niels Bohr Institute for support and
hospitality during his stays in Copenhagen
where part of this work was carried out.

\newpage
\ni {\Large\bf Figure Captions}
\vskip 1cm
{\bf Fig.\,1 }\ Feynman diagrams for the connected 4-point
$\pi \pi {\bar N}N$ Green function
$A_{\pi N}(q^2,q'^2)$. {\bf a)} Diagrams
with two external pion legs.
{\bf b)} and {\bf c)} Diagrams with one external pion leg.
The full lines represent the nucleon legs, the dashed lines the pion legs and
the double lines the external pseudo-scalar sources $j^a$.
\vskip 1cm
{\bf Fig.\,2 }\ The in-medium pion propagator.
The dashed line represents a free pion,
and the vertices $\odot$ and $\otimes$ are
given in Eq.(28). {\bf a)} The sum of all
contributions.\\ {\bf b)} The sum of the geometric series
involving $\odot$ contributions.


\newpage
\parindent 0 pt
\begin{thebibliography}{99}
\bibitem{kapnel}
D. B. Kaplan and A. E. Nelson, Phys. Lett. {\bf B175} (1986) 57.

\bibitem{mg}
A. Manohar and H. Georgi, Nucl. Phys. {\bf B234} (1984) 189;
H. Georgi, {\it Weak Interactions and Modern Particle Theory}
(~Benjamin/Cummings, Menlo Park, 1984~).

\bibitem{bkrt}
G. E. Brown, K. Kubodera, M. Rho and V. Thorsson, Phys. Lett.
{\bf B291} (1992) 355.

\bibitem{tpl}
V. Thorsson, M. Prakash and J. M. Lattimer,
Nucl. Phys. {\bf A572} (1994) 693; {\bf A574} (1994) 851.

\bibitem{blrt}
G. E. Brown, C.-H. Lee, M. Rho and V. Thorsson,
Nucl. Phys. {\bf A567} (1993) 937.

\bibitem{chlee}
C.-H. Lee, H. Jung, D.-P. Min and M. Rho, Phys. Lett.
{\bf B326} (1994) 14;
C.-H. Lee, G. E. Brown and M. Rho, SNUTP-94-28,
{\tt hep-ph/9403339};
C.-H. Lee, G. E. Brown, D.-P. Min and M. Rho,
SNUTP-94-50, {\tt hep-ph/9406311}.

\bibitem{ppt}
V. R. Pandharipande, C. J. Pethick and V. Thorsson,  work in progress.

\bibitem{bkr}
G. E. Brown, V. Koch and M. Rho, Nucl. Phys. {\bf A535} (1991) 701.

\bibitem{dee}
J. Delorme, M. Ericson and T.E.O. Ericson,
Phys. Lett. {\bf B291} (1992) 379.

\bibitem{sc1}
H. Yabu, S. Nakamura, F. Myhrer and K. Kubodera,
Phys. Lett. {\bf B315} (1993) 17; H. Yabu, S. Nakamura
and K. Kubodera, Phys. Lett. {\bf B317} (1993) 269.

\bibitem{sc2}
H. Yabu, F. Myhrer and K. Kubodera, Phys. Rev. {\bf D50} (1994) 3549.

\bibitem{ccwz}  S. Coleman, J. Wess and B. Zumino, Phys. Rev.
{\bf 177} (1969) 2239.

\bibitem{gl}
J. Gasser and H. Leutwyler, Ann. Phys. (N.Y.) {\bf 158} (1984) 142;
Nucl. Phys. {\bf B307} (1988) 763; (1991) 353.

\bibitem{gmor}
M. Gell-Mann, R. J. Oakes and B. Renner,
Phys. Rev. {\bf 175} (1968) 2195.

\bibitem{hohler}
G. H{\"o}hler, Landolt-B{\"o}rnstein, Vol. I/9b2, ed. H. Schopper
(~Springer, Berlin, 1983~).

\bibitem{chengdash}
T. P. Cheng and R. F. Dashen, Phys. Rev. Lett. {\bf 26} (1971) 594.

\bibitem{adldash}
See S. L. Adler and R. F. Dashen,
{\it Current algebra and applications to
particle physics} (~Benjamin, New York, 1968~).

\bibitem{gss}
J. Gasser, M. E. Sainio and A. \u{S}varc,
Nucl. Phys. {\bf B307} (1988) 779.

\bibitem{jenman}
E. Jenkins and A. Manohar, Phys. Lett. {\bf B255} (1991) 558.

\bibitem{bkm}
V. Bernard, N. Kaiser and U.-G. Mei{\ss}ner,
Phys. Lett. {\bf B309} (1993) 421.

\bibitem{gls}
J. Gasser, H. Leutwyler and M. E. Sainio,
Phys. Lett. {\bf B253} (1991) 252.

\bibitem{koch}
R. Koch, Nucl. Phys. {\bf A448} (1986) 707.

\bibitem{bkkm}
V. Bernard, N. Kaiser, J. Kambor and U.-G. Mei{\ss}ner,
Nucl. Phys.  {\bf B388} (1992) 315.

\bibitem{redef}
T. Appelquist and C. Bernard, Phys. Rev. {\bf D23} (1981) 425.

\bibitem{honerkamp}
J. Honerkamp, Nucl. Phys. {\bf B36} (1972) 130.

\bibitem{mannque}
C.-H. Lee and M. Rho, private communication.

\bibitem{polerefs}
G. Baym and E. Flowers, Nucl. Phys. {\bf A222}, (1974) 29;
A. E. Nelson and D. B. Kaplan, Phys. Lett. {\bf B192} (1987) 193;
T. Muto and T. Tatsumi, Phys. Lett. {\bf B283} (1992) 165.

\bibitem{pokorski}
S. Pokorski, {\it Gauge Field Theories}
(~Cambridge University Press, 1987~).

\bibitem{kirch}
M. Kirchbach and D. O. Riska, Nucl. Phys. {\bf A578} (1994) 511.

\bibitem{leutnonrel}
H. Leutwyler, Phys. Rev {\bf D49} (1994) 3033.

\bibitem{birse}
E. G. Drukarev and E. M. Levin, Nucl. Phys. {\bf A511} (1988) 697;
T. D. Cohen, R. J. Furnstahl and D. K. Griegel, Phys. Rev. Lett.
{\bf 67} (1991); Phys. Rev {\bf C45} (1992) 1881;
M. C. Birse, J. Phys. G. {\bf 20} (1994) 1537.

\bibitem{gmorrho}
V. Bernard and U.-G. Mei{\ss}ner Nucl. Phys. {\bf A489} (1988) 647;
M. Lutz, A. Steiner and W. Weise, Nucl. Phys. {\bf A542} (1992) 521;
{\bf A574} (1994) 755.

\bibitem{ks}
M. Knecht and J. Stern, IPNO-TH-94-53, to be published in
the second edition of the DAPHNE physics handbook,
Eds. L. Maini, G. Pancheri and N. Paver, {\tt hep-ph/9411253}.

\bibitem{wt}
A. Wirzba and V. Thorsson, to be published in the
proceedings of the workshop
{\em Hirschegg '95: Hadrons in Nuclear Matter}.

\bibitem{lambda}
M. J. Savage, Phys. Lett. {\bf B331} (1994) 411;
J. Schaffner, A. Gal, I. N. Mishustin, H. St{\"o}cker, and W. Greiner,
Phys. Lett. {\bf B334} (1994) 268;
V. Koch, Phys. Lett. {\bf B377} (1994) 7;
A. Steiner and W. Weise, to be published.

\bibitem{hyperons}
T. Muto, Prog. Theor. Phys. {\bf 89} (1993) 415;
P. J. Ellis, R. Knorren and M. Prakash, in progress.

\end{thebibliography}
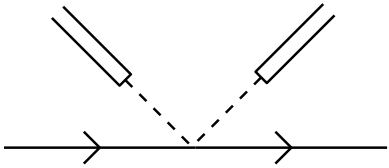
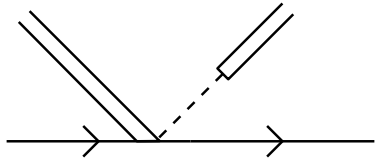
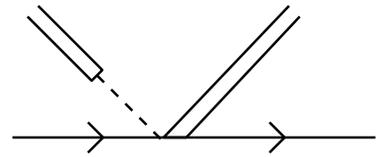
\end{document}